\documentstyle[aps,prl,twocolumn]{revtex}

\begin{document}

\twocolumn[\hsize\textwidth\columnwidth\hsize\csname
@twocolumnfalse\endcsname

\title{Analysis of soft optical modes in hexagonal BaTiO$_3$:\\
transference of perovskite local distortions}

\author{Jorge \'I\~niguez~\cite{email} and Alberto Garc\'{\i}a}
\address{
Departamento de F\'{\i}sica Aplicada II, Universidad del Pa\'{\i}s
Vasco, Apdo. 644, 48080 Bilbao, Spain}

\author{J.M. P\'erez-Mato}
\address{ 
Departamento de F\'{\i}sica de la Materia Condensada, 
Universidad del Pa\'{\i}s
Vasco, Apdo. 644, 48080 Bilbao, Spain}

\maketitle

\begin{abstract}
We have performed detailed first-principles calculations to determine
the eigenvectors of the zone-center modes of hexagonal BaTiO$_3$ and
shown that the experimentally relevant low-energy modes (including the
non-polar instability) can be represented as suitable combinations of
basic local polar distortions associated with the instability of the
cubic perovskite phase. The hexagonal structure provides a testing
ground for the analysis of the influence of the stacking of TiO$_6$
octahedra: the occurrence of relatively high-energy chains of dipoles
highlights the importance of local effects related to the coherent
hybridization enhancement between Ti and O ions. Our results provide
simple heuristic rules which could be useful for the analysis of
related compounds.
\end{abstract}



\vskip1pc
]

\narrowtext
\marginparwidth 2.7in
\marginparsep 0.5in

Barium titanate has two structural polymorphs: the cubic perovskite
type (c-BT) and its hexagonal modification (h-BT) with six formula
units per unit cell. While c-BT has been one of the best studied
ferroelectric materials for decades~\cite{lines-glass}, most of the
work on the structural and dielectric properties of h-BT is quite
recent~\cite{h-BT}. As shown in Fig.~\ref{fig1}, h-BT is also composed
of TiO$_6$ groups, albeit with a different stacking than the
perovskite form. It seems well established that the hexagonal
polymorph undergoes two zone-center structural phase transitions: at
$222$~K from the high temperature $P6_3/mmc$ hexagonal phase to a second
non-polar $C222_1$ phase, and at $74$~K to a ferroelectric $P2_1$
phase. The first transition is associated with the softening of an optical
mode and the second attributed to a shear strain instability; but a
detailed analysis is lacking due to the absence of structural
information on the two low-symmetry phases, and little is known about
the microscopic origin of the instabilities.

On the other hand, the discovery of a giant LO-TO splitting in h-BT by
Inoue {\sl et al.}~\cite{h-BT-pol} suggested that its ferroelectric
modes have the same origin as those of c-BT. In the cubic phase the
ferroelectric instabilities can be essentially described as chains of
dipoles that originate in the movement of Ti ions relative to their
surrounding O$_6$ octahedra with a minor distortion of the
later. Additional evidence in support of this view is provided by the
successful use of polar local modes by Zhong {\it et al.}~\cite{zvr}
in the construction of an effective Hamiltonian for c-BT.
In view of the basic structural similarities between the cubic and
hexagonal forms of BaTiO$_3$, it is meaningful to ask whether the
local modes that describe the unstable branches in c-BT can somehow be
transferred to h-BT and serve as a basis to discuss the low-energy
distortions of the structure.

Here we show the results of first-principles calculations that provide
for the first time structural information on the low-symmetry phases
of the hexagonal polymorph of BaTiO$_3$ and reveal the microscopic
nature of the modes. Our analysis proves that the structure of the
experimentally found zone-center optical soft~\cite{soft} modes in
h-BT is indeed characterized by the same distortions of the TiO$_6$
octahedra that are relevant in c-BT, leading to similar chains of
dipoles in the hexagonal structure.

In h-BT, the experimentally found optical soft modes are: the
zone-center instability that drives the phase transition at $222$~K
and transforms according to the $E_{2u}$ irreducible representation
(irrep) of $6/mmm$, and the $A_{2u}$ ferroelectric mode that softens
(though remaining stable) in the temperature range of the $C222_1$
phase and is responsible for the giant LO-TO splitting. Our
calculations agree with this experimental evidence. We performed a
full ab-initio~\cite{tech-det} relaxation of the thirty-atom h-BT
structure, resulting in lattice parameters $a$=$10.68$ and
$c$=$26.053$ a.u. (to be compared with experimental values of 10.77
and 26.451~\cite{exp-cell}, respectively). The five free internal
coordinates are also in excellent agreement with the experimentally
determined ones (within $1\%$).  After computing the force-constant
matrix at $\Gamma$ and diagonalizing it within the subspaces of the
appropriate symmetries, we found an unstable E$_{2u}$
mode~\cite{praga} and a soft but not unstable A$_{2u}$ mode. [It
should be noted that if the calculations are performed using the
experimental lattice parameters (i.e., at a larger cell volume), the
A$_{2u}$ mode is found to be unstable (and the E$_{2u}$ instability is
more pronounced). The fact that the A$_{2u}$ mode is very close to
being unstable could be particularly relevant for the phase transition
at $74$~K.]

A first analysis of the eigenvectors for both soft modes reveals that
the Ba contribution is small (around $10\%$ of the total mode norm,
compared to $4\%$ in the perovskite soft mode). Moreover, we checked
that if the Ba ions are frozen at their high-symmetry positions the
modes are still soft, so we do not consider them the following
discussion, and focus on the distortions of the TiO$_6$ groups.

To make the comparison to the perovskite quantitative, let us consider
the polar deformations of the TiO$_6$ groups of c-BT. These are shown
in Fig.~\ref{fig2}, where we assume that the Ti ion is located at the
origin of coordinates, so only the displacement patterns of the O ions
need to be considered. For each spatial direction $\alpha$=$x,y,z$ we
have two symmetry-adapted distortions denoted by $\hat{s}_{1,\alpha}$
and $\hat{s}_{2,\alpha}$ and transforming according to the $T_{1u}$
(vector like) irrep of $m3m$, the point group of the regular octahedra
of c-BT. In terms of this basis, the tetragonal ferroelectric
distortion in c-BT (along $x$ for concreteness) can be written as
$0.69\hat{s}_{1x}+0.73\hat{s}_{2x}$, with the distorted octahedra
exhibiting point group symmetry $4mm$~\cite{more-m3m}. In h-BT there
are two kinds of octahedra: those centered around Ti ions at $2a$
Wyckoff positions with $\bar{3}m$ point symmetry (denoted by
TiO$_6$(1) in Fig.~\ref{fig1}) and those arranged around Ti ions at
$4f$ Wyckoff positions with $3m$ point symmetry (TiO$_6$(2) in the
figure)~\cite{hex-polar}. The octahedra in the first set are
coordinated in the same way as in c-BT, i.e., by sharing O ions with
six other octahedra. Those in the second set are linked to other $1+3$
octahedra by sharing one O$_3$ face and single O ions respectively.

Due to the low (as compared to the case of c-BT) symmetry of the
octahedra in h-BT, their distortions associated to general $E_{2u}$
and $A_{2u}$ modes can be decomposed in a relatively large number of
symmetry-adapted displacement patterns. Among all the possible ones,
we restrict ourselves to those of c-BT type and check if they can
actually account for the structure of the soft
modes~\cite{we-can}. For instance, a general $A_{2u}$ distortion
leads the crystal to a phase with space group $P6_3mc$, in which the
TiO$_6$(1) groups reduce their point symmetry to $3m$ (see
Table~\ref{tab1}). As shown in Fig.~\ref{fig3}.b, c-BT distortions in
the $s_{ix}=s_{iy}=s_{iz}$ component combination (Ti ions move towards
O$_3$ faces as in the rhombohedric phase of c-BT) produce this
symmetry breaking. In the case of an $E_{2u}$ distortion, TiO$_6$(1)
reduces its point symmetry to $2$, and the appropriate c-BT mode has
the form $s_{ix}=-s_{iy}$ (orthorhombic), as shown in
Fig.~\ref{fig3}.c.

For the two soft modes of h-BT, we considered separately the various
classes of octahedra, computed from our ab-initio eigenvectors the
displacement of the O ions relative to the Ti ion, and performed a
projection of the resulting distortion field into the c-BT type
symmetry-adapted modes. The results (last column of Table~\ref{tab1})
present two main features: First, almost $100\%$ of the total
structural change associated with both soft modes can be described in
terms of the c-BT type polar distortions (normalization is chosen in
such a way that, for instance, for the first row in Table~\ref{tab1}
we have $(0.62^2+0.78^2)\times100=99.3\%$). Second, the components
$s_1;s_2$ are always similar in magnitude to those of c-BT
($0.69;0.73$) and present a positive $s_1/s_2$ ratio, which implies
that the O$_6$ octahedral cage moves almost rigidly relative to the Ti
ion also in h-BT. We have also computed the Born effective charge
associated with the ferroelectric $A_{2u}$ soft mode and found it
unusually large ($Z^*=11.29$), further confirming the relation with
the (rhombohedric) ferroelectric instability of c-BT (for which
$Z^*=9.956$)~\cite{z}.

We have proved then that at a local level the soft modes in h-BT can
be described by the same distortion vectors that determine the c-BT
polar instability.  In the crystal as a whole, these local polar
distortions lead to chains of dipoles, which points at the $E_{2u}$
instability and the softness of the $A_{2u}$ mode of h-BT as being
caused by Coulomb destabilizing forces, as it happens in the cubic
perovskite. The ferroelectric $A_{2u}$ soft mode, polarized along
$z'$, is roughly depicted in Fig.~\ref{fig1}. In the $E_{2u}$
distortion the chains of dipoles lay on the $x'y'$ plane and alternate
in orientation with a zero net polarization (the resulting $C222_1$
phase could be informally considered {\sl anti-ferroelectric} rather
than paraelectric).

From first-principles studies of the c-BT phase it is known that
parallel dipole chains are very weakly coupled, so that a transverse
modulation of a chain-like instability is not energetically relevant,
and, therefore, unstable TO normal modes exist almost in the whole
Brillouin Zone (BZ)~\cite{kx}. [The only exception are $k$ points near
${\bf k}_R = \frac{2\pi}{a}(1,1,1)$, for which we have an anti-phase
modulation of the Ti displacements (Ti$\Rightarrow$O$\Leftarrow$Ti--O)
in the three spatial directions, so the long-range destabilizing
forces are always canceled.] If this view is taken to its logical
conclusion, we could expect to find more zone-center soft modes in
h-BT, corresponding to the other possible distributions of chains of
dipoles. Table~\ref{tab2} enumerates all the possibilities. Apart from
the already discussed $A_{2u}$ and $E_{2u}$ modes, our ab-initio
calculations show that there is one $E_{1g}$ mode that is indeed
rather low in energy, while the ferroelectric $E_{1u}$ and the
$E_{2g}$ modes that are dominated by the movement of Ti ions are quite
hard. In order to explain this result, let us remark that for the
$E_{2u}$ and $A_{2u}$ soft modes the distortion is such that if an O
ion is approached by one of its two Ti neighbors the second Ti ion
moves away from it. This reflects the hybridization of the Ti $3d$ and
O $2p$ electronic states, which has been shown to be essential for the
occurrence of the c-BT ferroelectric
instability~\cite{cohen-ghosez}. It can be checked that any other
zone-center arrangement of the chains of dipoles results in either two
Ti ions approaching one O ion (for example, if the two Ti ions in one
of the O$_3-$Ti$-$O$_3-$Ti$-$O$_3$ groups depicted in Fig.~\ref{fig1}
move in the same way in the $x'y'$ plane, there is at least one oxygen
of the shared face that is approached by both) or in the second
titanium not moving away from an oxygen. In the former case ($E_{1u}$
and $E_{2g}$) the effect of the hybridization is lost and the
corresponding modes are hard. In the latter ($E_{1g}$ and $B_{1g}$),
the hardening is not as strong. Thus, we conclude that the particular
stacking of the TiO$_6$ groups in h-BT causes (through this local
effect) the relatively high energy of some chain-like distortions. We
can then formulate two ``rules of thumb'' for the characterization of
a given locally polar distortion as low-energy: First ($\cal R$1):
``There need to be chains of dipoles (without regard for their
transverse modulation)''. Second ($\cal R$2): ``The distribution of
such chains must lead to a {\sl coherent} hybridization enhancement
between Ti and O ions, where the word {\sl coherent} means that the
destabilizing effect is lost when two Ti ions approach the same O
ion''. These heuristic rules could be used to predict the occurrence
of locally polar soft modes at other $k$ points of the BZ of h-BT, as
well as in other structures with TiO$_6$ octahedra as basic building
blocks.

In summary, first-principles calculations of the character of the
zone-center modes of hexagonal BaTiO$_3$ support the physically
appealing idea that the experimentally relevant soft modes (including
the non-polar instability) can be represented as combinations of local
polar distortions transferred directly from the cubic perovskite form
of the compound. Our results lead also to heuristic rules that provide
insight into the influence of the arrangement of TiO$_6$ octahedra on
the low-energy dynamics of a structure~\cite{more-details}.

This work was supported in part by the UPV research grant
060.310-EA149/95 and by the Spanish Ministry of Education grant
PB97-0598. J.I. acknowledges financial support from the Basque
regional government.

\clearpage


\begin{table}
\begin{tabular}{lllll}
Mode  & Octahedra & Final  & $s_{i,\alpha}$ & Projection \\
      & type      & symmetry & distortion  & components \\
\tableline
$E_{2u}$ &TiO$_6$(1)$\bar{3}m$ & 2 & $s_x=-s_y$    & 0.62; 0.78 \\
         &TiO$_6$(2)$3m$       & 1 & $s_x,s_y,s_z$ & 0.05;0.01 ($x$) \\
         &                   &   &               & 0.44;0.50 ($y$) \\
         &                   &   &               & 0.51;0.53 ($z$) \\
         &                   &   &               & 0.68; 0.73 \\
\hline
$A_{2u}$ &TiO$_6$(1)$\bar{3}m$ & $3m$ & $s_x=s_y=s_z$ &   0.67; 0.70 \\
         &TiO$_6$(2)$3m$       & $3m$ & $s_x=s_y=s_z$ &   0.63; 0.74 \\
\end{tabular}
\vskip .5cm
\caption{Symmetry breakings of the h-BT phase TiO$_6$ groups
associated to the $A_{2u}$ and $E_{2u}$ soft modes. The fourth column
shows the combinations of symmetry-adapted c-BT type distortions that
are compatible with the symmetry reduction (it applies to both $s_1$
and $s_2$).
The last column shows the projections of the normalized total
distortion of the TiO$_6$ groups onto the $\hat{s}$ modes of the
second column, in the form $s_1;s_2$.
The $E_{2u}$ mode removes all symmetry elements from the TiO$_6$(2)
octahedra, and any combination of $s_{\alpha}$ is possible. For this
case we have
listed in the last column the $s_1$ and $s_2$ projections along each
of the three spatial directions as well as the modulus.}
\label{tab1}
\end{table}

\begin{table}
\begin{tabular}{lllllll}
Layer & $A_{2u}^f$ & $B_{1g}$ & $E_{1g}$ & $E_{1u}^f$ & $E_{2g}$ &
$E_{2u}$ \\
\tableline
Ti(2) & { }$\Uparrow$ & { }$\Uparrow$ & $\Rightarrow$ &
$\Rightarrow^*$ & $\Rightarrow^*$ & $\Rightarrow$ \\
Ti(2) & { }$\Uparrow$ & { }$\Uparrow$ & $\Leftarrow$ &
$\Rightarrow^*$ & $\Rightarrow^*$ & $\Leftarrow$ \\
Ti(1) & $\uparrow$ & & & $\rightarrow$ & & $\rightarrow$ \\
Ti(2) & { }{ }$\Uparrow$ & { }{ }$\Downarrow$ & $\Rightarrow$ &
$\Rightarrow^*$ & $\Leftarrow^*$ & $\Leftarrow$ \\
Ti(2) & { }{ }$\Uparrow$ & { }{ }$\Downarrow$ & $\Leftarrow$ &
$\Rightarrow^*$ & $\Leftarrow^*$ & $\Rightarrow$ \\
Ti(1) & $\uparrow$ & & & $\rightarrow$ & & $\leftarrow$ \\
\tableline
      & 0.0074  & ($\geq$ 0.046) &0.0105  &0.3305  &0.2543  &-0.0123\\
\end{tabular}
\vskip .5cm
\caption{Symbolic description of the possible zone-center chains of
dipoles in h-BT, classified in terms of irreps of $6/mmm$. The layers
of Ti ions are represented along the $z'$ direction as in
Fig.~\protect\ref{fig1} (Ti(1) and Ti(2) refer to Ti ions in
TiO$_6$(1) and TiO$_6$(2) groups respectively). Arrows indicate the
orientation of the dipoles (horizontal ones symbolize any direction in
the $x'y'$ plane), and those set in the same type are symmetry
related. The superscript $f$ marks the ferroelectric
modes. Displacement patterns that violate $\cal R$2 (see text) are
marked with an asterisk. The bottom line shows the mode force
constants in atomic units (for $B_{1g}$ an unambiguous assignment
cannot be made).}
\label{tab2}
\end{table}

\begin{figure}
\caption{Unit cell of the $P6_3/mmc$ phase of h-BT. For the sake of
simplicity, only Ti and O ions are shown. The distributions of local
dipoles corresponding to the $E_{2u}$ and $A_{2u}$ soft modes are
indicated.}
\label{fig1}
\end{figure}

\begin{figure}
\caption{Polar (vector like) deformations of the $m3m$ (regular)
octahedra of c-BT. Only the modes polarized along $x$ are
indicated. The $y$ and $z$ sets are analogous.}
\label{fig2}
\end{figure}

\begin{figure}
\caption{Combinations of $\hat{s}_{i,\alpha}$ modes that represent the
symmetry breaking of the octahedra in h-BT by the soft modes. Panel
a)~shows a conveniently oriented octahedron which will be assumed to
have $\bar{3}m$ (resp. $3m$) symmetry. Oxygen ions are labeled as in
Fig.~\protect\ref{fig2} and the cartesian axes with origin in the Ti
ion are indicated. Panel b)~shows the $s_x=s_y=s_z$ distortion that
breaks the 2-fold axes (resp. no symmetry element) and is related to
the $A_{2u}$ mode. In panel c), a $s_x=-s_y$ distortion results in a 2
(resp. 1) point-symmetry, as appropriate for the $E_{2u}$ mode. Only
$\hat{s}_1$ modes are shown, but the same combinations apply to
$\hat{s}_2$.}
\label{fig3}
\end{figure}


\begin{references}
\bibitem{soft} We use the term {\sl soft} to refer to instable modes
as well as low-energy (though maybe stable) modes.
\bibitem{tech-det} We have used the local-density approximation and
the pseudopotential approach with Vanderbilt's ultrasoft
pseudopotentials. The electron wave functions were expanded in a plane
wave basis with an energy cut-off of 25 Rydberg, and the BZ sums were
calculated by a $3\times3\times(2(+0.5))$ Monkhorst-Pack special
k-point mesh.
\bibitem{more-m3m} A rhombohedral distortion ($3m$) would be
represented by equal amplitudes of the three components of $\hat{s}_1$
and $\hat{s}_2$ (maintaining the same $(0.69,0.73)$ mix), and an
orthorhombic one ($222$) by equal amplitudes for two of the components
and zero for the third.
\bibitem{hex-polar} In this case, the octahedra exhibit a polar
distortion even in the $P6_3/mmc$ phase of h-BT: both Ti ions in a
O$_3-$Ti$-$O$_3-$Ti$-$O$_3$ set are displaced along $z'$ towards the
O$_3$ faces on the side (which agrees with the $\cal R$2 heuristic
rule in the text),
producing two opposite local dipoles and no net polarization.
\bibitem{we-can} It can be proven (by group-theoretical
considerations) that polar distortions located at both $2a$ and $4f$
Wyckoff positions produce zone-center modes of the desired symmetries.
\bibitem{more-details} A more detailed account of the work presented
here, as well as other results (such as the existence of a previously
unreported zone-center soft Rigid Unit Mode) will be presented
elsewhere.
\end{references}
\end{document}